\documentclass[a4paper]{jpconf}
\usepackage{amssymb}
\usepackage{amsthm}

\usepackage{latexsym}
\newcommand{\be}{\begin{eqnarray}}
\newcommand{\ee}{\end{eqnarray}}
\newcommand{\bdm}{\begin{displaymath}}
\newcommand{\edm}{\end{displaymath}}

\newcommand{\gth}[1]{\textsf{T}_\textsf{#1}}
\def\SIM{\triangleq}
\begin{document}          
\title{\textbf{Solution generating theorems for perfect fluid spheres}}
\author{\textbf{Petarpa Boonserm, Matt Visser, and Silke Weinfurtner}}
\date{}
\address{School of Mathematics, Statistics, and Computer Science,\\ 
Victoria University of Wellington, PO Box 600, Wellington, New Zealand\\}
\ead{\textbf{Petarpa.Boonserm@mcs.vuw.ac.nz, matt.visser@mcs.vuw.ac.nz, and silke.weinfurtner@mcs.vuw.ac.nz}}
 \vspace{1cm}
\begin{abstract}
The first static spherically symmetric perfect fluid solution with constant density was found by Schwarzschild in 1918. Generically, perfect fluid spheres are interesting because they are first approximations to any attempt at building a realistic model for a general relativistic star. Over the past 90 years a confusing tangle of specific perfect fluid spheres has been discovered, with most of these examples seemingly independent from each other. To bring some order to this collection, we develop several new transformation theorems that map perfect fluid spheres into perfect fluid spheres. These transformation theorems sometimes lead to unexpected connections between previously known perfect fluid spheres, sometimes lead to new previously unknown perfect fluid spheres, and in general can be used to develop a systematic way of classifying the set of all perfect fluid spheres. In addition, we develop new ``solution generating'' theorems for the TOV, whereby any given solution can be ``deformed'' to a new solution. Because these TOV-based theorems work directly in terms of the pressure profile and density profile it is relatively easy to impose regularity conditions at the centre of the fluid sphere.
 \end{abstract}

\section{Introduction}
Perfect fluid spheres, either Newtonian or relativistic, are the first approximations in developing realistic stellar models~\cite{Delgaty, Skea, exact}.
Because of the importance of these models,
fully general ``algorithmic'' solutions of the perfect fluid constraint in general relativity have been developed over the last several years~\cite{Rahman, Lake, Martin, Petarpa, Petarpa1}.  For some of the key highlights of earlier history regarding perfect fluid spheres see references~\cite{Buchdahl, Bondi, Wyman, Hojman-et-al}, while for  a recent analysis providing upper and lower bounds on the internal structure of a relativistic perfect fluid sphere one might consult~\cite{Martin0}. 

For our current purposes, the central idea is to start solely with spherical symmetry, which implies that in orthonormal components the stress energy tensor takes the form:
\begin{equation}
T_{\hat a\hat b} = \left[ \begin{array}{cccc}
\rho&0&0&0\\ 0&p_r&0&0\\ 0&0&p_t&0\\ 0&0&0&p_t \end{array}\right],
\end{equation}
and then use the perfect fluid constraint $p_r=p_t$. This simply makes the radial pressure equal to the transverse pressure.
By using the Einstein equations, plus spherical symmetry, the equality  $p_r=p_t$ for the pressures becomes the statement
\begin{equation}
G_{\hat\theta\hat\theta} = G_{\hat r\hat r} = G_{\hat\phi\hat\phi}.
\end{equation}
In terms of the metric components, this leads to an ordinary differential equation [ODE], which then constrains the spacetime geometry, for \emph{any} perfect fluid sphere.

The big recent change has been the discovery and development of ``algorithmic" techniques that permit one to generate large classes of perfect fluid spheres in a purely mechanical way~\cite{Rahman, Lake, Martin, Petarpa, Petarpa1}. In this article we will summarize and discuss our recent work on  these algorithmic ideas~\cite{Petarpa, Petarpa1}, providing several solution-generating theorems of varying levels of complexity.  We shall then explore the formal properties of these solution-generating theorems, and shall then use these theorems to classify some of the previously known exact solutions. Moreover, we shall then generate several new previously unknown perfect fluid solutions, and finally we shall show how one can rephrase all these theorems directly in terms of the pressure profile and density proflie.

\section{Solution generating theorems}
Start with some static spherically symmetric geometry in Schwarzschild (curvature) coordinates
\begin{equation} \label{line_element_1}
ds^2 = - \zeta(r)^2 \; dt^2 + {dr^2\over B(r)} + r^2 \;d\Omega^2,
\end{equation}
and assume it represents a perfect fluid sphere. That is,
$ G_{\hat\theta\hat\theta} = G_{\hat r\hat r} =G_{\hat\phi\hat\phi} $.
While $G_{\hat\theta\hat\theta} = G_{\hat\phi\hat\phi}$ is always fulfilled due to spherical symmetry,
setting $G_{\hat r\hat r} = G_{\hat\theta\hat\theta}$ supplies us with an ODE
\begin{equation} 
\label{ode_for_B}
[r(r\zeta)']B'+[2r^2\zeta''-2(r\zeta)']B + 2\zeta=0 \, ,
\end{equation}
which reduces the freedom to choose the two functions in equation (\ref{line_element_1})
to one. This equation is a first order-linear non-homogeneous equation in $B(r)$. Thus
--- once you have chosen a $\zeta(r)$ --- this equation can always be solved for $B(r)$.
Solving for $B(r)$ in terms of $\zeta(r)$ is the basis of the analyses in references ~\cite{Lake,Martin}, (and is the basis for Theorem 1 below).
On the other hand, we can also re-group this same equation as
\begin{equation}    
\label{ode_for_zeta}
2 r^2 B \zeta'' + (r^2 B'-2rB) \zeta' +(r B'-2B+2)\zeta=0 \,,
\end{equation}
which is a linear homogeneous 2nd order ODE for $\zeta(r)$, which will become the basis for Theorem 2 below.
The question we are going to answer in this section is, how to systematically ``deform'' this geometry
while still retaining the perfect fluid property.
That is, suppose we start with the specific geometry defined by
\begin{equation}
ds^2 = - \zeta_0(r)^2 \; dt^2 + {dr^2\over B_0(r)} + r^2 d\Omega^2
\end{equation}
and assume it represents a perfect fluid sphere.
We will show how to ``deform'' this solution by applying four different transformation theorems on $\left\{ \zeta_0 , B_0  \right\}$,
such that the outcome still presents a perfect fluid sphere. The outcome of this process will depend on one or more free parameters, and so automatically generates an entire family of perfect fluid spheres of which the original starting point is only one member. We also try to find the connection between all different transformation theorems

\subsection{Four  theorems}
The first theorem we present is a variant of a result  first explicitly published in reference \cite{Martin}.

\paragraph{Theorem 1} 
Suppose $\{ \zeta_0(r), B_0(r) \}$ represents a perfect fluid sphere.
Define
\begin{equation} \label{Theorem1_m_1}
\Delta_0(r)  =
 \left({ \zeta_0(r)\over  \zeta_0(r) + r  \;\zeta'_0(r)}\right)^2 \; r^2 \; 
\exp\left\{ 2 \int {\zeta'_0(r)\over  \zeta_0(r)} \; 
  { \zeta_0(r)- r\; \zeta'_0(r)\over  \zeta_0(r) + r  \;\zeta'_0(r)} \; d r\right\}.
\end{equation}
Then for all $\lambda$, the geometry defined by holding $\zeta_0(r)$ fixed and
setting
\begin{equation}
ds^2 = - \zeta_0(r)^2 \; dt^2 + {dr^2\over B_0(r)+\lambda\; \Delta_0(r) }
+ r^2 d\Omega^2
\end{equation}
is also a perfect fluid sphere. That is, the mapping
\begin{equation}
\gth{1}(\lambda): \left\{ \zeta_0 , B_0  \right\} \mapsto 
\left\{ \zeta_0 , B_0 + \lambda\Delta_0(\zeta_0) \right\}
\end{equation}
takes perfect fluid spheres into perfect fluid spheres. 

\paragraph{Proof for Theorem 1}
Assume that $\left\{ \zeta_{0}(r),B_{0}(r) \right\}$ is a solution for equation (\ref{ode_for_B}).
Under what conditions does $\left\{ \zeta_{0}(r),B_1(r) \right\}$  also satisfy equation (\ref{ode_for_B})?
Without loss of generality, we write
\begin{equation}
B_1(r)=B_{0}(r) + \lambda\;\Delta_{0}(r) \, .
\end{equation}
Equation ({\ref{ode_for_B}}) can now be used
to determine $\Delta_{0}(r)$.
That ordinary \emph{inhomogeneous} first-order differential equation in $B$ now
simplifies to
\begin{equation}
\label{ode_th1}
\left[ r (r \zeta_0 )' \right] \Delta_{0}'
+ \left[ 2 r^2 \zeta_{0}'' - 2 (r \zeta_{0})' \right] \Delta_{0} = 0   \, ,
\end{equation}
which is an ordinary \emph{homogeneous} first-order differential equation in $\Delta_{0}$.
A straightforward calculation, including an integration by parts, leads to
\begin{equation}
\Delta_0(r) = \frac{r^2 }{\left[ (r \zeta_{0})' \right]^2}  \; \exp\left\{{\int{\frac{4 \zeta_{0}'}{(r \zeta_{0})'} dr } }\right\} \, .
\end{equation}
Adding and subtracting 
$\pm 2 (r \zeta_{0}'^2)/(\zeta_{0} (r \zeta_{0})')$ to the argument under the integral leads to
\begin{equation}
\Delta_{0}
= \left({ \zeta_0(r)\over  \zeta_0(r) + r  \;\zeta'_0(r)}\right)^2 \; r^2 \; 
\exp\left\{ 2 \int {\zeta'_0(r)\over  \zeta_0(r)} \; 
  { \zeta_0(r)- r\; \zeta'_0(r)\over  \zeta_0(r) + r  \;\zeta'_0(r)} \; d r\right\} \, ,
\end{equation}
as advertised.

\paragraph{Comments:}
A version of Theorem 1  can also be found in~\cite{exact}. 
Specifically, after several manipulations, changes of notation, and a change of coordinate system, the transformation exhibited in equation (16.11) of~\cite{exact} can be  converted into Theorem 1 above.

\paragraph{Theorem 2} 
Let $\{\zeta_0,B_0\}$ describe a perfect fluid sphere.
Define
\begin{equation}
Z_0(r) = \sigma +\epsilon \int {r \; dr\over  \zeta_0(r)^2\; \sqrt{B_0(r)} }.
\end{equation}
Then for all $\sigma$ and $\epsilon$, the geometry defined by holding $B_0(r)$ fixed and
setting
\begin{equation}
ds^2 = - \zeta_0(r)^2 \; Z_0(r)^2
\; dt^2 + {dr^2\over B_0(r)} + r^2 d\Omega^2
\end{equation}
is also a perfect fluid sphere.
That is, the mapping
\begin{equation}
\gth{2}(\sigma,\epsilon): \left\{ \zeta_0, B_0  \right\} \mapsto \left\{ \zeta_0 \; Z_0(\zeta_0,B_0), B_0 \right\}
\end{equation}
takes perfect fluid spheres into perfect fluid spheres. 

\paragraph{Proof for Theorem 2}
The proof of theorem 2 is based on the technique of ``reduction in order''. 
Assuming that $\left\{ \zeta_{0}(r),B_{0}(r) \right\}$  solves equation (\ref{ode_for_zeta}), 
write 
\begin{equation}
\zeta_1(r)=\zeta_{0}(r) \; Z_0(r) \, .
\end{equation}
and demand that $\left\{ \zeta_{1}(r),B_{0}(r) \right\}$ also solves equation (\ref{ode_for_zeta}).
We find
\begin{equation}
\label{ode_th2}
(r^2 \zeta_0 B_0' + 4 r^2 \zeta_0' B_0 - 2 r \zeta_0 B_0) Z_0' + (2 r^2 \zeta_0 B_0) Z_0''  =0 \, ,
\end{equation}
which is
an ordinary homogeneous second-order differential equation, depending only on $Z_0'$
and $Z_0''$. (So it can be viewed as a first-order homogeneous order differential equation in $Z'$, which is solvable.)
Separating the unknown variable to one side,
\begin{equation} \label{de_for_zetaprime}
\frac{Z_0''}{Z_0'}=-\frac{1}{2} \frac{B_0'}{B_0} - 2 \frac{\zeta_0'}{\zeta_0} + \frac{1}{r} \, .
\end{equation}
Re-write $Z_0''/ Z_0' = d\ln(Z_0')/d t$, and integrate twice over both sides of
equation (\ref{de_for_zetaprime}), to obtain
\begin{equation} \label{eq_for_zeta_1}
Z_0=\sigma + \epsilon \int{\frac{r \, dr}{\zeta_0(r)^2 \;\sqrt{B_0(r)} }}  \, ,
\end{equation} 
depending on the old solution $\left\{ \zeta_0 (r) , B_0 (r)  \right\}$, and two
arbitrary integration constants $\sigma$ and $\epsilon$.  

\bigskip

Having now found the first and second generating theorems makes it possible to define two new theorems by composing them.
Take a perfect fluid sphere solution $\left\{ \zeta_0, B_0  \right\}$. Applying Theorem 1 onto it gives
us a new perfect fluid sphere $\left\{ \zeta_0, B_1 \right\}$. The new $B_1$ is given in
equation (\ref{Theorem1_m_1}). If we now continuing with applying Theorem 2, again we get a
new solution $\{ \tilde\zeta, B_1 \}$, where $\tilde{\zeta}$ now depends on the new $B_1$.
All together we can consider this as a single process, by introducing the following theorem: \\

\paragraph{Theorem 3}
If $\{\zeta_0,B_0\}$ denotes a perfect fluid sphere, then for all $\sigma$,  $\epsilon$, and $\lambda$, the three-parameter geometry defined by
\begin{equation}
ds^2 = - \zeta_0(r)^2 
\left\{
\sigma+\epsilon \int {r \; dr\over  \zeta_0(r)^2\; \sqrt{B_0(r)+\lambda \;\Delta_0(r)}} 
\right\}^2
\; dt^2 + {dr^2\over B_0(r) +\lambda \Delta_0(r)   }
+ r^2 d\Omega^2
\end{equation}
is also a perfect fluid sphere, where $\Delta_0$ is 
\begin{equation}
\Delta_0(r) =
\left({ \zeta_0(r)\over  \zeta_0(r) + r  \;\zeta'_0(r)}\right)^2 \; r^2 \; 
\exp\left\{ 2 \int {\zeta'_0(r)\over  \zeta_0(r)} \; 
  { \zeta_0(r)- r\; \zeta'_0(r)\over  \zeta_0(r) + r  \;\zeta'_0(r)} \; dr\right\} \, .
\end{equation}
That is
\begin{equation}
\gth{3} = \gth{2}\circ \gth{1}: \{\zeta_0,B_0\} \mapsto \{\zeta_0,B_0+\lambda\;\Delta_0(\zeta_0)\} 
\mapsto \{\zeta_0 \; Z_0(\zeta_0,B_0+\lambda\Delta_0(\zeta_0)),B_0+\lambda\Delta_0(\zeta_0)\}
\end{equation}
takes perfect fluid spheres into perfect fluid spheres.

\bigskip
Now, instead of starting with theorem 1 we could first apply theorem 2 on $\left\{ \zeta_0, B_0  \right\}$.
This gives us a new perfect fluid sphere $\left\{ \zeta_1, B_0  \right\}$, 
where $\zeta_1=\zeta_0 \; Z_0(\zeta_0,B_0)$ is given by equation (\ref{eq_for_zeta_1}). Continuing with
theorem 1 leads to $\{ \zeta_1, \tilde{B}  \}$, where $\tilde{B}$ depends on the new
$\zeta_1$. Again, we can consider this as a single process, by introducing the following theorem:\\

\paragraph{Theorem 4}
If $\{\zeta_0,B_0\}$ denotes a perfect fluid sphere, then for all $\sigma$,  $\epsilon$, and $\lambda$, the three-parameter geometry defined by
\begin{equation}
ds^2 = - \zeta_0(r)^2 
\left\{
\sigma+\epsilon \int {r \; dr\over  \zeta_0(r)^2\; \sqrt{ B_0(r)   }} 
\right\}^2
\; dt^2 + {dr^2\over B_0(r) + \lambda \Delta_0(\zeta_1,r)  }
+ r^2 d\Omega^2
\end{equation}
is also a perfect fluid sphere, where $\Delta_0(\zeta_1,r)$ is defined as
\begin{equation}
\Delta_0(\zeta_1,r)=
\left({ \zeta_1(r)\over  \zeta_1(r) + r  \;\zeta_1'(r)}\right)^2 \; r^2 \; 
\exp\left\{ 2 \int {\zeta_1'(r)\over  \zeta_1(r)} \; 
  { \zeta_1(r)- r\; \zeta_1'(r)\over  \zeta_1(r) + r  \;\zeta'_1(r)} \; dr\right\} \, ,
\end{equation}
depending on $\zeta_1 = \zeta_0\;  Z_0$, where as before
\begin{equation}
Z_0(r) = 
\sigma+\epsilon \int {r \; dr\over  \zeta_0(r)^2\; \sqrt{ B_0(r) }}  \, .
\end{equation}
That is
\begin{equation}
\gth{4} = \gth{1}\circ \gth{2}: \{\zeta_0,B_0\} \mapsto \{\zeta_0 \;Z_0(\zeta_0,B_0),B_0\} 
\mapsto \{\zeta_0 \; Z_0(\zeta_0,B_0),B_0+\lambda \Delta_0(\zeta_0 \; Z_0(\zeta_0,B_0))\}
\end{equation}
takes perfect fluid spheres into perfect fluid spheres.

\bigskip
Below we present two tables which describe the inter-relationships of certain examples of perfect fluid spheres under $\gth{1}$, $\gth{2}$, $\gth{3}$, and $\gth{4}$.
We define a seed metric as one for which Theorem 1 and Theorem 2 both yield output metrics distinct from the input metric: $\gth{1}(g) \not\SIM g \not\SIM \gth{2}(g)$. In contrast for non-seed metrics one or the other of these theorems is trivial, either $\gth{1}(g)\SIM g$ or $\gth{2}(g)\SIM g$ (where $g$ is the metric or a parameterized class of metrics). Classifying perfect fluid spheres into seed and non-seed metrics is a very useful step. Our theorems are effectively a device to generate new solutions for a perfect fluid sphere without \emph{explicitly} solving the Einstein equations. (Of course our theorems are set up in such a way that one is \emph{implicitly} solving the Einstein equations.)  The terminology used in the tables is more fully explained in reference~\cite{Petarpa}.

\begin{table}[!ht]
\centerline{Some seed geometries and their descendants.}
\bigskip
\begin{tabular}{|  l  |  l  |  l  |  l  | l |}
\hline
Seed & Theorem1 & Theorem2 & Theorem 3 & Theorem 4\\
\hline
Minkowski & Einstein static & K-O III & int Schwarzschild & Martin 3\\ 
ext Schwarzschild & Kuch68 II & Kuch 86 I  &    [integral] & [integral] \\
de Sitter& Tolman IV   & int Schwarzschild  &  P7 & [integral] \\
Tolman V ($A=0$) & Tolman V &  Tolman VI & Wyman III & Wyman IIa\\
S1 & Tolman V ($n=+1$) & K--O III & P4 & Martin 3 \\
M--W III & Martin 2 &  P5 &  P6 & [integral] \\
Heint IIa (C=0) &  Heint IIa & P3 & P8 & [integral] \\
\hline
\hline
\end{tabular}
\caption{Seed solutions and their generalizations derived via theorems 1--4. The notation ``[integral]'' denotes a metric so complicated that explicitly writing out the relevant integral is so tedious that it does not seem worthwhile. The terminology is more fully explained, and the relevant metrics explicitly presented, in reference~\cite{Petarpa}.}
\bigskip
\end{table}

\begin{table}[!ht]
\centerline{Some non-seed perfect fluid geometries and their descendants.}
\bigskip
\hskip 2.2 cm
\begin{tabular}{|  l  |  l  |  l  |  l  | l |}
\hline
Base & Theorem1 & Theorem2 & Theorem 3 & Theorem 4\\
\hline
Tolman IV & Tolman IV &   P7  &   P7  & [integral] \\
B-L & B-L &Kuchb I b & Kuchb I b  & [integral]  \\
Heint IIa &  Heint IIa & P8 & P8 & [integral] \\
\hline

Tolman VI& Wyman IIa &  Tolman VI & [integral] & [integral] \\

Kuch1 Ib & Martin 1 & Kuch1 Ib &  P2 & Martin 1\\

K--O III & Martin 3&  K--O III &   P1  & Martin 3\\
\hline



\hline
\end{tabular}
\caption{Non-seed solutions and their generalizations. The terminology used in this table is more fully explained in reference~\cite{Petarpa}.}
\bigskip
\end{table}

\section{Solution generating theorems for the TOV equation}
The Tolman--Oppenheimer--Volkov [TOV] equation constrains the internal structure of general relativistic static perfect fluid spheres \cite{Petarpa1}. We have developed several ``solution generating'' theorems for the TOV, whereby any given solution can be deformed to a new solution. Because these theorems work directly in terms of the physical observables --- the pressure profile and density profile --- it is relatively easy to examine the density and pressure profiles for physical reasonableness. This section complements our previous theorems wherein a similar ``algorithmic'' analysis of the general relativistic static perfect fluid sphere was presented directly in terms of the spacetime geometry --- in the present analysis the pressure and density are primary and the spacetime geometry is secondary.
Specifically, consider the well-known TOV system of equations, whose
derivation is now a common textbook
exercise~\cite{Wald,Misner}:
\begin{eqnarray}
{dp(r)\over dr} &=&
- { [\rho(r)+p(r)]\;[m(r)+ 4\pi p(r)\, r^3]
\over 
r^2 [1-2m(r)/r]};
\\
{dm(r)\over dr} &=& 4\pi \,\rho(r) \, r^2.
\end{eqnarray}
We adopt units where $G=c=1$ so that $m/r$ and $p\, r^2$ are dimensionless.
Our basic strategy will be to assume that somehow we have obtained, or
been given, some specific ``physically reasonable'' solution of the TOV
equation in terms of the profiles $p_0(r)$ and $\rho_0(r)$.


Viewed as a differential equation for $p(r)$, with $m(r)$ [and hence
$\rho(r)$] specified, the TOV equation is a specific example of a
Ricatti equation, for which the number of useful solution techniques
is unfortunately rather limited.
Using standard results (see the explicit discussion in reference~\cite{Petarpa1}, or for example the general references~\cite{Bender, Polyanin}) it is relatively simple to present
the following:

\paragraph{Theorem {\bf P1}}
  Let $p_0(r)$ and $\rho_0(r)$ solve the TOV equation and hold $m_0(r)
  = 4\pi \int \rho_0(r) \, r^2 \, dr$ fixed. Define
\begin{equation}
g_0(r) = {m_0(r)+4\pi p_0(r) \,r^3\over r^2[1-2m_0(r)/r]}.
\end{equation}
Then the general solution to the TOV equation is $p(r) = p_0(r)+\delta
p(r)$ where
\begin{equation}
\delta p(r) =
{
{\displaystyle
\delta p_c \;\sqrt{1-2m_0/r} \; \exp\left\{ - 2 \int_{0}^r g_0 \,dr \right\} 
}
\over
{\displaystyle
1 + 4\pi \delta p_c \int_{0}^r {\displaystyle{1\over \sqrt{1-2m_0/r}}}
\exp\left\{ -2\int_{0}^r g_0 \,dr \right\} r \, dr
}
},
\end{equation}
and where $\delta p_c$ is the shift in the central pressure.

A second theorem can be obtained by looking for \emph{correlated}
changes in the mass and pressure profiles:

\paragraph{Theorem {\bf P2}}
  Let $p_0(r)$ and $\rho_0(r)$ solve the TOV equation and hold $g_0$
  fixed, in the sense that
\begin{equation}
g_0(r) = {m_0(r)+4\pi p_0(r) r^3\over r^2[1-2m_0(r)/r]} =  {m(r)+4\pi p(r) r^3\over r^2[1-2m(r)/r]}.
\end{equation}
Then the general solution to the TOV equation is given by $p(r) =
p_0(r)+\delta p(r)$ and $m(r) = m_0(r)+\delta m(r)$ where
\begin{equation}
\delta m(r) =   {4\pi r^3 \; \delta \rho_c  \over 3} \;{1\over  [1+r\;g_0]^2} \; 
\exp\left\{  2 \int_0^r g_0 \; {1-r\,g_0\over 1+r\,g_0} dr \right\}.
\end{equation}
and 
\begin{equation}
\delta p(r) =  - { \delta m\over 4\pi r^3} \; {1+8\pi p_0 r^2 \over 1-2m_0/r}.
\end{equation}
Here $\delta \rho_c$ is the shift in the central density.

\section{Discussion}

Using Schwarzschild coordinates we have developed two fundamental transformation theorems that map perfect fluid spheres into perfect fluid spheres. Moreover, we established two additional transformation theorems by composing the first and second generating theorems. We then used these transformations as a basis for classifying different types of perfect fluid sphere solutions.
If we apply these theorems on a known perfect fluid sphere, different
solutions are often obtained. While some of the solutions are already known in the literature, most of them are novel.
Therefore, we have developed a tool to generate new solutions for a perfect fluid sphere, but which does not
require us to directly solve the Einstein equations. In principle \emph{all} solutions of the perfect fluid constraint should be obtainable in this manner.
In short, perfect fluid spheres may ``simple'', but they still provide a surprisingly rich mathematical and physical structure.

Furthermore, we have also developed two ``physically clean'' solution-generating theorems for the TOV equation --- where by
``physically clean'' we mean that it is relatively easy to understand
what happens to the pressure and density profiles, especially in the
vicinity of the stellar core.

Finally, we reiterate that the relativistic perfect fluid sphere,
despite its venerable history, continues to provide interesting
surprises.

\ack This research was supported by the Marsden Fund administered by the Royal Society of New Zealand. In addition, PB was supported by a Royal Thai Scholarship and a Victoria University Small Research Grant. SW was supported by the Marsden Fund, by a Victoria University PhD Completion Scholarship, and a Victoria University Small Research Grant.

\end{document}